\documentclass[preprint,12pt, a4paper]{elsarticle}

\usepackage{amssymb}
\usepackage{hyperref}
\usepackage{array}
\journal{SoftwareX}

\begin{document}
\renewcommand{\labelenumii}{\arabic{enumi}.\arabic{enumii}}

\begin{frontmatter}



\title{CompFlowLab: A Python code to develop and prototype new data-driven models for challenging compressible flow problems with shocks and chemical reactions}


\author[ku]{Ali Mohaghegh\corref{cor1}}
\ead{ali.mohaghegh@ku.edu}
\author[ku]{Cheng Huang}
\ead{chenghuang@ku.edu}
\cortext[cor1]{Corresponding author}
\address[ku]{University of Kansas, Aerospace Engineering, Lawrence, KS 66045, USA}

\begin{abstract}
CompFlowLab is an open-source Python environment that is capable of modeling different classes of compressible flow problems (including shocks, flames, and detonation waves) using a one-dimensional (1D) compressible Navier-Stokes solver with multi-species transport and chemical reaction models. It is designed specifically for the data-driven modeling community as a lightweight, accessible prototyping platform to test, develop, and evaluate new modeling methods on numerically and physically challenging compressible flow problems, especially those featuring shocks and chemical reactions. Specifically, CompFlowLab aims at providing computationally efficient calculations on convection-dominanted problems that are well recognized to be difficult for conventional data-driven modeling techniques, such as shocks, flames, and detonation waves, and more importantly enabling rapid testing and prototyping of new data-driven models. The code serves three primary purposes: (1) generating high-fidelity full order model data for training of data-driven modeling, (2) providing a modular platform for implementing and testing novel data-driven algorithms (e.g., machine learning methods and reduced order modeling techniques) on challenging physics, and (3) offering a standalone Computational Fluid Dynamics (CFD) solver with validated test cases that can also support numerical method development in the broader CFD community. By unifying these capabilities in a clean, extensible Python codebase, CompFlowLab lowers barriers to innovation at the intersection of model reduction and complex fluid dynamics.

\end{abstract}

\begin{keyword}
Computational Fluid Dynamics (CFD) \sep Compressible Flow \sep Reduced Order Modeling \sep Data-Driven Modeling \sep Combustion



\end{keyword}

\end{frontmatter}


\section*{Required Metadata}
\label{}

\begin{table}[!h]
\begin{tabular}{|l|p{6.5cm}|p{6.5cm}|}
\hline
\textbf{Nr.} & \textbf{Code metadata description} & \textbf{Please fill in this column} \\
\hline
C1 & Current code version & V2.2.12 \\
\hline
C2 & Permanent GitHub link to code/repository used for this code version & \href{https://github.com/alimike97/CompFlowLab}{https://github.com/alimike97/CompFlowLab}\\
\hline
C3 & Legal Code License   & MIT \\
\hline
C4 & Code versioning system used & git \\
\hline
C5 & Software code languages, tools, and services used & Python 3; NumPy, SciPy, scikit-learn, Matplotlib, Cantera; pytest; hatchling (packaging); GitHub Actions (CI) \\
\hline
C6 & Compilation requirements, operating environments \& dependencies & No compilation. Python $\geq$ 3.12, NumPy, SciPy, scikit-learn, Matplotlib, Cantera $\geq$ 3.2. Linux, macOS, Windows. Install with pip install compflowlab.\\
\hline
C7 & If available Link to developer documentation/manual & \href{https://compflowlab.mintlify.app}{https://compflowlab.mintlify.app} \\
\hline
C8 & Support email for questions & mohaghegh.ali76@gmail.com\\
\hline
\end{tabular}
\caption{Code metadata (mandatory)}
\label{} 
\end{table}

\newpage

\section{Motivation and significance}
\label{sec:motivation}

Data-driven models, from machine-learning surrogates to projection-based
reduced order models (ROMs), offer substantial potential for accelerating computationally intensive many-query tasks
such as design, uncertainty quantification, and control
\cite{lumley_low-dimensional_1997,benner_survey_2015,mohaghegh_machine_2025}.
However, constructing accurate and efficient ROMs remains particularly challenging for
convection-dominated flows with shocks, flames, and detonations.
The transport of sharp features across the domain often leads to slow convergence of the approximation error with increasing linear-subspace dimension, a fundamental limitation known as the
Kolmogorov $n$-width barrier
\cite{cohen_optimal_2020, bonnaillie-noel_efficient_2016, peherstorfer_breaking_2022, klein_multi-fidelity_2026}.
To address this challenge, recent remedies have explored nonlinear and quadratic manifolds
\cite{lee_model_2020, barnett_quadratic_2022, geelen_operator_2023}, online adaptation of basis and
sample points
\cite{peherstorfer_online_2015, huang_predictive_2023, hedayat_-context_2026, hedayat_toward_2026,mohaghegh_feature-guided_2026}. Developing and systematically evaluating these methods requires a computational testbed that captures the essential interactions among shocks, chemical reactions, and
reactive transport, while preserving sufficiently efficient and accessible to support rapid algorithm development, testing and validation. Specifically, we identify four major gaps that hinder the translation of fundamental methodological advances in data-driven modeling into practical applications involving challenging physical problems:

\begin{enumerate}
  \item \label{lim:model}
  \textbf{The model-problem gap:} Many research codes and tutorials for
  data-driven techniques development and demonstration rely on simplified benchmark problems, such as the Burgers', heat, or linear advection equations.
  While these problems enable rapid prototyping and preliminary validation, they do not adequately represent the strong nonlinearities and the multi-physics interactions in highly compressible-flow problems. Consequently, methods
  that perform well on these simplified problems may not retain their accuracy, robustness, or computational efficiency when applied to more realistic flows.

  \item \label{lim:data}
  \textbf{The data generation gap:} Existing libraries, including PyROM \cite{puzyrev_pyrom_2019},
  RBniCS \cite{rozza_real_2024}, and packages for proper orthogonal decomposition (POD), dynamic mode decomposition (DMD)
  \cite{demo_pydmd_2018}, and operator inference
  \cite{peherstorfer_data-driven_2016}, provide reliable tools for constructing
  data-driven models. General-purpose model reduction libraries such as pyMOR
  \cite{milk_pymor_2016}, libROM \cite{choi_librom_2019}, and Pressio
  \cite{rizzi_pressio_2020} supply additional machinery for ROM development and deployment. However, these tools usually rely on external full order model (FOM) solvers for generating training data or require users to supply and integrate their own high-fidelity solvers. For complex compressible flows, configuring, validating, and executing such solvers can demand substantial expertise and computational resources, creating a significant barrier to developing and testing new data-driven methodologies.

  \item \label{lim:access}
  \textbf{The accessibility and computational-cost gap:} High-fidelity simulations of realistic compressible flows often rely on large,
  specialized CFD codes, some of which are proprietary or have restricted accessibility. Even when the source code is available, simulations may require substantial high-performance computing resources and long runtime, limiting their suitability for rapid algorithm development and testing. Moreover, intrusive ROM formulations, such as Galerkin and least-squares
  Petrov--Galerkin (LSPG) projection \cite{carlberg_galerkin_2017}, hyper-reduction \cite{everson_karhunenloeve_1995}, and adaptive ROMs \cite{mohaghegh_feature-guided_2026, hedayat_history-aware_2026}, require direct access to the governing-equation residuals. Though open-source CFD frameworks such as OpenFOAM \cite{jasak_openfoam_2009} and SU2 \cite{economon_su2_2016} provide access to their numerical implementations, extracting and modifying the necessary solver operations can require substantial effort due to the complexity of their software architectures. On the other hand, developing a dedicated high-fidelity solver from scratch presents an alternative, but the associated implementation and validation effort can divert considerable resources from the primary objective of advancing data-driven modeling techniques.

  \item \label{lim:compare}
  \textbf{The consistent-comparison gap:} Conventional projection-based ROMs, nonlinear-manifold ROMs, and adaptive ROMs are frequently developed and evaluated using different in-house solvers with different numerical treatments, physical models, and benchmark configurations. As a result, reported differences in accuracy, stability, and computational performance may reflect not only the underlying modeling methodologies but also differences in numerical implementation and simulation setup. Furthermore, intrusive and non-intrusive approaches are typically not evaluated under identifical computational configurations. Therefore, such lack of a common, physically representative benchmark environment complicates systematic comparisons and makes it difficult to isolate the relative advantages and limitations of different data-driven modeling strategies.
\end{enumerate}

CompFlowLab is therefore developed to address these gaps by providing an open-source, Python-based computational environment that integrates a CFD solver for the one-dimensional (1D) compressible Navier--Stokes equations with multi-species transport and chemical reactions into a modular data-driven modeling framework. The software is designed to serve three primary purposes: (1) generate high-fidelity FOM solutions and training datasets for data-driven model development; (2) provide an accessible and extensible platform for implementing, testing, and evaluating advanced data-driven methodologies, including machine-learning methods and projection-based ROMs, on challenging physically representative compressible-flow problems; and (3) offer a standalone CFD solver with validated benchmark cases to support broader research and development for CFD community in physical models and numerical methods.

These capabilities directly address the four gaps identified above. First, built-in benchmark problems involving shocks, flames, and detonations
provide physically representative test cases for evaluating data-driven methods beyond simplified models (limitation~\ref{lim:model}). Second, the integrated, validated finite volume FOM that enables direct generation of high-fidelity training data within the same computational environment, eliminating the need to configure and couple an external CFD solver (limitation~\ref{lim:data}). Third, the modular implementation exposes the governing equation operators and supports residuals evaluations at arbitrarily selected computational cells, which facilitates the development of intrusive ROMs and hyper-reduction techniques without extensive modifications to a complex CFD solver. In addition, this lightweight implementation allows representative benchmark simulations on a single CPU core, thus substantially lowering the accessibility and computational barriers for algorithm development (limitation~\ref{lim:access}). Finally, a unified simulation interface supports both FOM simulations and nine ROM solver modes using consistent governing equations and numerical treatments, enabling systematic comparisons of different modeling strategies while minimizing discrepancies arising from solver implementations (limitation~\ref{lim:compare}).

We remark that the development of CompFlowLab is also motivated by the tradeoffs associated with extending existing software infrastructure versus creating a dedicated research environment. Incorporating a
compressible-flow solver into an existing general-purpose data-driven modeling library would require substantial redevelopment and integration effort while adapting a production CFD code for rapid methodological prototyping would require navigating and modifying a considerably morecomplex software infrastructure. CompFlowLab provides a purpose-built, lightweight environment in which the FOM solver and the data-driven modeling components are clearly separated, modular, and readily accessible for modification and extension. This design prioritizes methodological flexibility, computational efficiency, and ease of experimentation while retaining the essential physics needed to evaluate advanced data-driven modeling techniques for compressible flows. Specifically, we compare CompFlowLab with the closely related existing tools in Table~\ref{tab:related} .

\begin{table}[!h]
\begin{tabular}{|>{\raggedright\arraybackslash}p{2.7cm}|>{\raggedright\arraybackslash}p{1.7cm}|>{\raggedright\arraybackslash}p{1.8cm}|>{\raggedright\arraybackslash}p{1.6cm}|>{\raggedright\arraybackslash}p{1.7cm}|>{\raggedright\arraybackslash}p{1.6cm}|}
\hline
\textbf{Tool} & \textbf{Language} & \textbf{Built-in compressible FOM} & \textbf{Detailed chemistry} & \textbf{Intrusive ROM with hyper-reduction} & \textbf{Online adaptive ROM} \\
\hline
CompFlowLab & Python & Yes & Yes (Cantera) & Yes & Yes \\
\hline
PERFORM \cite{wentland_perform_2022} & Python & Yes & No & Yes (No hyper-reduction) & No \\
\hline
PyROM \cite{puzyrev_pyrom_2019} & Python & No & No & No & No \\
\hline
RBniCS \cite{rozza_real_2024} & Python (FEniCS) & Yes & No & Yes & No \\
\hline
Pressio \cite{rizzi_pressio_2020} & C{+}{+} / Python & Euler test problems & No & Yes & No \\
\hline
libROM \cite{choi_librom_2019} & C{+}{+} & No (couples to external codes) & No & Yes & No \\
\hline
pyMOR \cite{milk_pymor_2016} & Python & No (bindings to external solvers) & No & Yes & No \\
\hline
PyDMD \cite{demo_pydmd_2018} & Python & No & No & No (non-intrusive) & No \\
\hline
Operator Inference \cite{peherstorfer_data-driven_2016} & Python & No & No & No (non-intrusive) & No \\
\hline
\end{tabular}
\caption{Comparison with related open-source tools.}
\label{tab:related}
\end{table}

This paper presents two primary contributions: \textbf{(1)} we introduce CompFlowLab that, to the best of our knowledge, is the first open-source Python-based software package integrating a compressible-flow solver with modeling capabilities handing detailed chemical kinetics with multiple ROM methodologies, including conventional projection-based, adaptive, and nonlinear-manifold-based approaches; \textbf{(2)} we establish and demonstrate a well-documented reproducible workflow encompassing FOM data generation, ROM construction, and performance evaluation across a suite of benchmark problems ranging from inert shock dynamics to 1D rotating detonation engine (RDE).

A key feature of CompFlowLab is its streamlined user interface, which enables different modeling approaches to be deployed with minimal changes to the simulation setup. Users can specify the problem configuration and solver options through a single plain-text input file containing
\texttt{keyword = value} pairs and executes \texttt{python main.py <working\_dir> <input\_file>}. In FOM mode, the solver outputs conserved variables, primitive variables, and residuals as NumPy arrays at a user-defined interval of \texttt{save\_interval} steps. To construct and execute a conventional projection-based ROM, users can set \texttt{solver\_mode = PROM} and specifying the location of the FOM training snapshots via \texttt{training\_data\_dir}. The software then constructs a proper
orthogonal decomposition (POD) basis according to a prescribed energy threshold and advances the
ROM using the same numerical setups. Alternatively, setting \texttt{solver\_mode = AROM} activates the adaptive ROM, which initializes the basis using a short window of FOM simulations and subsequently updates the basis and sample points during online execution, eliminating the need for separate offline training stage.


\section{Software description}
\label{sec:software}

CompFlowLab is an open-source, Python-based software package consisting of approximately 7,600 lines of code, organized into eight subpackages under \texttt{src/compflowlab} (Table~\ref{tab:package}). It
depends only on a limited set of established libraries, including NumPy \cite{harris_array_2020}, SciPy
\cite{virtanen_scipy_2020}, scikit-learn \cite{pedregosa_scikit-learn_2011},
Matplotlib \cite{hunter_matplotlib_2007}, and Cantera \cite{speth_canteracantera_2017},
and is distributed on the Python Package Index (PyPI) (i.e., straightforward installation using \texttt{pip install compflowlab}). Computationally intensive operations, such as numerical flux evaluation, time stepping, and basis update, are implemented using vectorized NumPy operations, enabling code execution on a single CPU core without compilation. The software architecture is guided by three primary design principles: (1) \emph{accessibility and simplicity} that provide a lightweight, easy to access environment for rapid prototyping and evaluation of data-driven modeling techniques; (2) \emph{modularity and extensibility} that maintain a modular, clear separation between the FOM and the data-driven modeling components to facilitate independent development and integration of new algorithms; and (3) \emph{reliability and reproducibility} that leverage reliable established libraries numerical operations and physical modeling. Specifically, it uses Cantera for standardized thermodynamic properties, transport models, and chemical kinetics, while relying on NumPy and SciPy for efficient numerical computations without compromising code readability or ease of modification. Together, these principles establish a flexible computational environment that balances physical modeling capabilities, computational efficiency, and accessibility for research and algorithm development.

\subsection{Software architecture}
\label{sec:architecture}

\begin{figure}[!h]
\centering
\includegraphics[width=\textwidth]{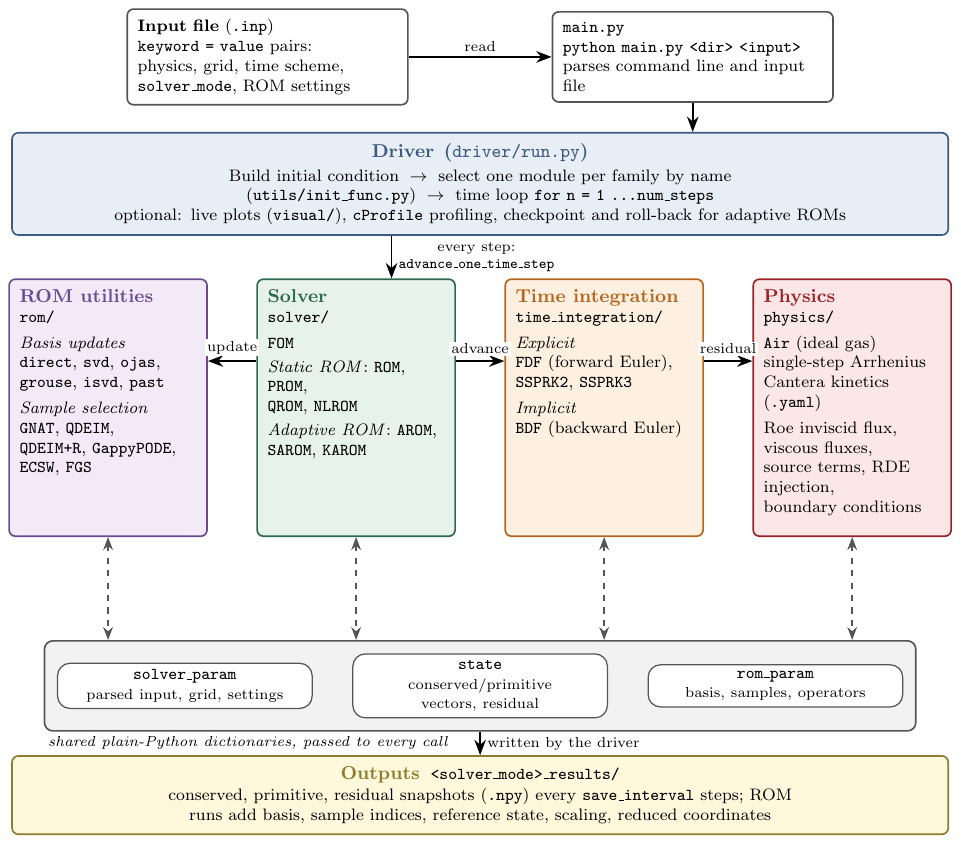}
\caption{CompFlowLab architecture. The driver selects one module from each family by name from the input file and passes three dictionaries between them; any solver module that implements \texttt{advance\_one\_time\_step} can be plugged in.}
\label{fig:architecture}
\end{figure}

The CompFlowLab architecture is built around a unified data interface in which the software components communicate through three Python dictionaries as illustrated inFig.~\ref{fig:architecture}. The \texttt{solver\_param} dictionary stores user-defined parsed simulation parameters from the input file, together with derived quantities such as the grid spacing. The \texttt{state} dictionary contains the conserved and primitive state vectors, governing-equation residual, and cached thermodynamic properties. The \texttt{rom\_param} dictionary stores ROM-specific information, including basis, reference state, scaling vectors, sample indices, and precomputed hyper-reduction operators. This dictionary-based design provides direct access to intermediate quantities without requiring users to navigate complex class hierarchies, facilitating debugging, inspection, and modification during algorithm development.

\begin{sloppypar}
The simulation workflow consists of four primary stages. First, \texttt{main.py} processes the command line arguments and parses the user-specified input file. Second, \texttt{driver/run.py} establishes the initial condition and selects the appropriate physics module, a time integrator, and a solver according to the input configuration. Third, the advances through a unified time-stepping interface:
\begin{center}
\small\texttt{solver.advance\_one\_time\_step(solver\_param, state, physics, time\_integration, rom\_param)}
\end{center}
Finally, simulation outputs are saved to the \texttt{<solver\_mode>\_results} directory within the working directory. Additional capabilities, including live Matplotlib visualization of selected flow variables and performance profiling via \texttt{cProfile}, are integrated into the same execution workflow. For adaptive ROM simulations, the software also provides a checkpoint that restores a previous simulation state and extends the FOM simulation window when a ROM time step fails, thereby ensuring robustness during simulations.
\end{sloppypar}

\begin{sloppypar}
The software modularity and extensibility are achieved through three standardized module interfaces: (1) The physics module provides the \texttt{residual\_calculator} functions based on the selected gas model (ideal air or reacting flow); (2) The time-integration module implements \texttt{advance\_time} while each solver module implements \texttt{advance\_one\_time\_step}. These common interfaces allow new modeling approaches to be incorporate with minimal changes to the existing software. For example, implementing a new ROM function simply requires adding a solver
module under \texttt{solver/} and registering it through \texttt{init\_func.init\_solver}. The same modular structure also supports composite modeling strategies that combine existing formulations. In particular, the self-adaptive ROM leverages the adaptive ROM during transient regimes and subsequently transitions to a static ROM when the evolving dynamics can be adequately represented without further basis adaptation.
\end{sloppypar}

The underlying software infrastructure are also designed to support hyper-reduction, which requires evaluating the governing-equation residuals only at selected computational locations. The full state is stored as a flatten, variable-major vector, including two ghost cells at each domain boundary. Utility functions in \texttt{utils/reshape\_func.py} provide
conversions between the solver layout and a variables-by-cells arrangement, as well as consistent mappings of sampling indices between the two layouts. When hyper-reduction is enabled, the physics modules evaluate numerical fluxes only at the faces bounding sampled cells, which avoids full-domain residual evaluations and makes the cost of a residual scaling with the number of samples rather than the number of cells, thereby providing the computational savings expected by hyper-reduction. The software package structure of Compflowlab is summerized in Table~\ref{tab:package}. 

\begin{table}[!h]
\begin{tabular}{|l|>{\raggedright\arraybackslash}p{4.2cm}|>{\raggedright\arraybackslash}p{4.3cm}|}
\hline
\textbf{Subpackage} & \textbf{Files} & \textbf{Role} \\
\hline
\texttt{driver} & \texttt{run.py} & Initialization and the main time integration loop \\
\hline
\texttt{physics} & \texttt{Ideal\_Air.py}, \newline \texttt{Single\_Step
\_Reacting\_Flow.py}, \texttt{Reacting\_Flow.py} & State conversions, Roe inviscid and viscous fluxes, chemical source terms, RDE injection \\
\hline
\texttt{time\_integration} & \texttt{FDF.py}, \texttt{SSPRK2.py}, \texttt{SSPRK3.py}, \texttt{BDF.py} & Explicit and implicit time integrators, including LSPG \\
\hline
\texttt{solver} & \texttt{FOM.py}, \texttt{ROM.py}, \texttt{PROM.py}, \texttt{QROM.py}, \texttt{NLROM.py}, \texttt{AROM.py}, \texttt{SAROM.py}, \texttt{KAROM.py}& FOM and seven ROM solver modes, each implementing \texttt{advance\_one\_time\_step} \\
\hline
\texttt{rom} & \texttt{basis\_func.py}, \texttt{sampling\_func.py} & Online basis-update rules and hyper-reduction sample selection \\
\hline
\texttt{boundary\_condition} & \texttt{bc\_func.py} & Ghost-cell updates \\
\hline
\texttt{utils} & \texttt{input\_read\_func.py}, \texttt{init\_func.py}, \texttt{reshape\_func.py} & Input parsing, module selection, data layout, snapshot assembly \\
\hline
\texttt{visual} & \texttt{visual\_func.py} & Live plotting \\
\hline
\end{tabular}
\caption{Software package structure of CompFlowLab (\texttt{src/compflowlab}).}
\label{tab:package}
\end{table}

\subsection{Software functionalities}
\label{sec:functionalities}

\emph{Full order model:} CompFlowLab solves the one-dimensional compressible Navier--Stokes equations in conservative form
\begin{equation}
    \frac{\partial \mathbf{q}}{\partial t} + \frac{\mathbf{\partial f}}{\partial x}  - \frac{\mathbf{\partial f_{v}}}{\partial x} = \mathbf{H}
    ,
  \qquad
  \mathbf{q} = \left[\rho,\ \rho u,\ \rho E,\ \rho Y_1,\ \dots,\ \rho Y_{N_{s-1}}\right]^{T},
  \label{eq:ns}
\end{equation}
where $\mathbf{q}$ is the vector of conserved variables, $\mathbf{f}$ and $\mathbf{f}_v$ are the inviscid and viscous fluxes and $\mathbf{H}$ represents the chemical source terms. Here, $\rho$, $u$, $E$, and $Y_i$ denote the density, velocity, total specific energy, and mass fraction of species $i$, respectively, while $N_s$ is the total number of chemical species. The mass fraction of the final species is determined from the constraint $\sum_{i=1}^{N_s}Y_i=1$. Spatial discretization is achieved using a cell-centered finite-volume method on a uniform computational grid with inviscid interface fluxes evaluated using the Roe approximate Riemann solver \cite{roe_approximate_1981}. For reacting mixtures, the Roe-averaged thermodynamic and transport properties are evaluated through Cantera. Viscous, heat conduction, and species diffusion fluxes are included when \texttt{viscous = True}. Three gas-model options are supported through the \texttt{gas\_model} input parameter. The \texttt{Air} option represents a calorically perfect gas with a constant specific heat ratio of $\gamma = 1.4$, requiring only three conserved variables. Alternatively, specifying a \texttt{.inp} file activates a two species,
single step Arrhenius reaction model with model constants (pre-exponential factor, activation energy, heat capacity, molecular weight, enthalpies of formation) read directly from the input file. For reacting-flow simulations with realistic chemistry, a Cantera \texttt{.yaml} mechanism file can be supplied to incorporate detailed chemical kinetics and the associated thermochemical properties - e.g., the ten-species hydrogen/oxygen mechanism provided in \texttt{h2o2.yaml}. Boundary conditions are specified independently for each primitive variable at both domain boundaries with available options including periodic, reflecting wall, extrapolation, and prescribed-value conditions. In addition, a simplified reactant-injection model is implemented to represent the continuous supply of fresh reactants in 1D RDE simulations.

Four time integration, CompFlowLab provides four schemes: forward Euler (\texttt{FDF}), the strong-stability-preserving Runge--Kutta schemes SSPRK2 and SSPRK3 \cite{durran_numerical_2010}, and backward Euler (\texttt{BDF}) \cite{kelley_solving_2003, knoll_jacobian-free_2004, baker_technique_2005}. All four schemes share a common interface and support both full-order and hyper-reduced state representations, enabling consistent temporal integration across the different solver formulations.

\begin{table}[!h]
\begin{tabular}{|l|>{\raggedright\arraybackslash}p{9.0cm}|>{\raggedright\arraybackslash}p{1.0cm}|}
\hline
\textbf{\texttt{solver\_mode}} & \textbf{Method} & \textbf{Ref} \\
\hline
\texttt{ROM} & Static POD basis; the full order update is computed at sampled cells and the full state is reconstructed each step through the gappy POD operator $\mathbf{V}(\mathbf{S}^{T}\mathbf{V})^{+}$ & \cite{everson_karhunenloeve_1995} \\
\hline
\texttt{PROM} & Projection ROM with Galerkin (explicit) or LSPG (implicit) projection, optionally hyper-reduced & \cite{carlberg_galerkin_2017} \\
\hline
\texttt{QROM} & Quadratic manifold: a regularized least-squares quadratic correction added to the linear POD approximation & \cite{geelen_operator_2023, barnett_quadratic_2022} \\
\hline
\texttt{NLROM} & POD with a learned closure that predicts truncated coordinates from retained ones (gaussian process, or radial basis function regression) & \cite{ares_de_parga_nonlinear_2026} \\
\hline
\texttt{AROM} & Adaptive ROM: an initial FOM window, then hyper-reduced steps with an online basis update every step and periodic sample updates & \cite{peherstorfer_data-driven_2016, huang_predictive_2023, mohaghegh_feature-guided_2026} \\
\hline
\texttt{SAROM} & Self-adaptive ROM: runs \texttt{AROM}, detects recurring solution patterns, gathers data with an incremental SVD, then switches to a cheaper static ROM & \cite{mohaghegh_self_2026, mohaghegh_self-adaptive_2026} \\
\hline
\texttt{KAROM} & \texttt{AROM} variant for prototyping incremental-SVD and subspace tracking basis updates & \cite{hedayat_history-aware_2026} \\
\hline
\end{tabular}
\caption{ROM solver modes, selected with the \texttt{solver\_mode} keyword.}
\label{tab:rom_modes}
\end{table}

\emph{Reduced-order models:} CompFlowLab provides seven ROM solver modes that share the same governing-equation operators
and time-integration schemes as the FOM, thus enabling consistent implementation and comparison of different ROM strategies (Table~\ref{tab:rom_modes}). ROM is constructed using POD of centered and variable normalized FOM snapshots, with the number of POD modes determined by a user-specified energy threshold (\texttt{pod\_energy}). Alternatively, precomputed bases can be loaded directly, eliminating the need for repeated basis construction.

To support flexible ROM development, CompFlowLab provides multiple interchangeable hyper-reduction and basis-adaptation algorithms that can be combined with different compatible ROM formulations. Specifically, six hyper-reduction sampling strategies are available through the parameter
\texttt{hyper\_method}: \texttt{GNAT} \cite{carlberg_corrigendum_2013}, \texttt{QDEIM}
  \cite{drmac_new_2015}, \texttt{QDEIM+R} \cite{wentland_investigation_2021}, \texttt{GappyPODE} \cite{peherstorfer_stability_2020}, \texttt{ECSW} \cite{farhat_dimensional_2014}, \texttt{FGS} \cite{mohaghegh_feature-guided_2026}. The number of sampling points is controlled by \texttt{sampling\_rate},expressed as a percentage of the computational cells. For adaptive ROM, six basis-adaptation methods are available through the parameter \texttt{arom\_method}: \texttt{direct} \cite{mohaghegh_feature-guided_2026}, a least-squares update over a rolling snapshot window \cite{peherstorfer_online_2015}; \texttt{ojas}, Oja's rule \cite{oja_simplified_1982}; \texttt{grouse}, a rank one geodesic step on the Grassmann manifold \cite{balzano_online_2010}; \texttt{isvd}, an incremental SVD with a forgetting factor \cite{goos_incremental_2002,brand_fast_2006}; and \texttt{past}, projection-approximation subspace tracking \cite{bin_yang_projection_1995}. These interchangeable components facilitate systematic investigations of the effects of sampling and basis-adaptation strategies on ROM accuracy, stability, and computational efficiency.

\emph{Outputs and utilities:} CompFlowLab provides standardized simulation outputs to facilitate post-processing, reproducibility, and quantitative comparisons across solver configurations. Each simulation saves and outputs conservative and primitive variables, governing-equation residuals, and heat release rates for reacting-flow cases in NumPy \texttt{.npy} format. In addition, ROM simulations store the basis, sample indices, reference state, scaling vectors, and reduced coordinates, which allows intermediate quantities to be examined after the simulation. Software reliability is supported by an automated \texttt{pytest} suite comprising 48 unit tests covering the physics modules, time integrators, ROM utilities, and input handling. These tests are executed through GitHub Actions on every code push to provide continuous verification of core software functionality.

\section{Illustrative examples}
\label{sec:examples}

CompFlowLab provides a suite of ready-to-run benchmark cases ranging from non-reacting shock problems, laminar flames, and detonations, the corresponding input configurations of which are summarized in Table~\ref{tab:cases}. The non-reacting cases feature analytical or well documented reference solutions and can be completed within seconds on a standard laptop, enabling rapid verification and testing of numerical methods. The reacting cases incorporate stiff chemical kinetics and strongly coupled multi-scale dynamics, which provide more challenging test cases for evaluating the performance of new data-driven modeling approaches.

\begin{table}[!h]
\begin{tabular}{|>{\raggedright\arraybackslash}p{5cm}|>{\raggedright\arraybackslash}p{7.75cm}|}
\hline
\textbf{Case} & \textbf{Main difficulty for ROMs} \\
\hline
Sod shock tube \cite{sod_survey_1978} & Shock, contact discontinuity, and rarefaction \\
\hline
Double diaphragm shock tube & Interacting waves from two discontinuities \\
\hline
Colliding shocks & Head on collision of strong shocks \\
\hline
Sedov blast wave \cite{ganapa_blast_2021} & Strong shocks and near vacuum core \\
\hline
Free standing flame & Stiff chemistry and thin reaction zone \\
\hline
Transient and perturbed flames & Flame front traveling across the grid \\
\hline
Wall reflected detonation & Shock reflection, ignition delay, and transition to detonation \cite{oran_weak_1982}\\
\hline
Rotating detonation engine (1D) \cite{mohaghegh_self_2026}& Self sustained traveling detonation with continuous injection \\
\hline
\end{tabular}
\caption{Bundled test cases in \texttt{examples} directory}
\label{tab:cases}
\end{table}

\emph{Worked example on the Sod shock tube from FOM to adaptive ROM:}
To illustrate the CompFlowLab workflow, we consider the Sod shock-tube problem and demonstrate how a FOM simulation can be extended to conventional ROM and adaptive ROM simulations through minimal modifications to the input configuration. The computational domain $[0, 1]$~m is discretized into 500 uniformly spaced cells. The solution is advanced for 5,000 forward Euler time steps with a fixed step size $\Delta t = 5 \times 10^{-5}$~s to a final time of $t = 0.25$~s. The relevant parameters in the input file are:

{\small
\begin{verbatim}
solver_mode       = FOM          # step 1; then PROM, then AROM
dt                = 5e-5
num_steps         = 5000
time_scheme       = FDF
cell_number       = 500
x_interval_ic     = [[0,0.5],[0.5,1]]
rho_ic            = [1,0.125]
press_ic          = [1,0.1]
gas_model         = Air
flux_scheme       = 2nd Order Roe
save_interval     = 10
# ROM settings (ignored in FOM mode)
training_data_dir = ./FOM_results
pod_energy        = 99.999
rom_method        = galerkin
hyper             = True
hyper_method      = QDEIM
sampling_rate     = 10           # percent of cells
arom_method       = direct
init_training_win = 10
unsampled_update_freq = 10
\end{verbatim}
}

\noindent All three simulations are executed using the same command:

{\small
\begin{verbatim}
python main.py examples/sod_shock_tube examples/sod_shock_tube/input_file.inp
\end{verbatim}
}

\begin{sloppypar}
\noindent In step~1, the FOM simulation generates and outputs 500 snapshots, saved under the \texttt{FOM\_results} directory for subsequent ROM construction and evaluation. In
step~2, conventional projection-based ROM is activated through \texttt{solver\_mode = PROM}, which (1) constructs a POD basis from these saved snapshots by retaining 99.999\% of the total POD energy, (2) selects sampling locations for hyper-reduction using QDEIM, and (3) advances the resulting Galerkin ROM in time. In step~3, adaptive ROM is activated by setting
\texttt{solver\_mode = AROM}, which requires no offline training stage. Instead, the simulation begins with a short initialization window of 10 FOM time steps and then advances via a hyper-reduced ROM which updates its basis every step and refreshes its sampling points every 10 steps. Moreover, the modular configuration of CompFlowLab also enables alternative sampling strategies to be evaluated with minimal effort. For example, switching from \texttt{QDEIM} to \texttt{FGS} requires changing only the \texttt{hyper\_method} parameter.
\end{sloppypar}

Beyond the Sod shock-tube example, CompFlowLab provides additional test cases listed in Table~\ref{tab:cases}. Details of all bundled test cases, including setup configurations, together with representative simulation results, screenshots, and videos  are available in the online documentation at \url{https://compflowlab.mintlify.app/test-cases}.

\section{Impact}
\label{sec:impact}

CompFlowLab provides an accessible and computationally efficient platform for advancing data-driven modeling methodologies for compressible flows. By bridging the gap between simplified benchmark problems and computationally demanding production-scale CFD solvers, it enables researchers to develop and evaluate new algorithms on physically representative problems involving shocks, flames, and detonations without the substantial effort typically required to implement, configure, and validate a dedicated CFD solver. Its modular architecture and unified simulation interface facilitate several important research applications:

\begin{enumerate}

  \item \textbf{Rapid prototyping of new methodologies:} Researchers
  can conveniently implement and evaluate new data-driven modeling techniques, such as projection formulations, sampling strategies, basis-update algorithms, and learned closure models, by modifying or extending individual module. The integrated
  FOM solver, data-generation pipeline, and benchmark configurations allow researchers to focus on methodological innovations rather than the development of supporting software infrastructure, which expedite the new algorithm development across a range of representative compressible flow problems.
  
  \item \textbf{Systematic, reproducible, and consistent comparisons:} Different types of ROM methods, including the conventional projection-based ROM, nonlinear-manifold ROM and adaptive ROM, can be evaluated using consistent governing equations, spatial discretizations, numerical fluxes, and time-integration schemes. This common computational environment minimizes discrepancies arising from differences in solver implementations and enables more direct assessments of the accuracy, stability, and computational efficiency of competing modeling approaches. Furthermore, interchangeable hyper-reduction and basis-adaptation algorithms facilitate controlled investigations of individual algorithmic components.

  \item \textbf{Evaluation of intrusive and non-intrusive modeling approaches:} Direct access to the governing-equation residuals and their evaluation at selected computational locations enables the development of intrusive ROM formulations, including projection-based, hyper-reduced, and adaptive approaches. At the same time, the integrated FOM data-generation capabilities support non-intrusive methodologies that learn reduced dynamics or surrogate representations directly from simulation data. Both classes of methods can therefore be developed and evaluated using common training datasets and benchmark configurations, facilitating systematic investigations of their relative advantages and limitations.
  
  \item \textbf{Generation of training and validation datasets:} The FOM provides time-resolved flow variables, governing-equation residuals, and relevant thermochemical quantities for compressible problems. These datasets can support the development and validation of scientific machine-learning models, including neural-network surrogates, learned reduced dynamics, nonlinear-manifold representations, and data-driven closure models. By eliminating the need for external production-scale CFD software, CompFlowLab reduces the computational and implementation barriers to generating representative datasets for methodological research.
\end{enumerate}

\emph{Demonstrated research applications:} CompFlowLab has already supported the development and evaluation of several data-driven modeling techniques for challenging compressible flows. These include feature-guided sampling strategies \cite{mohaghegh_feature-guided_2026}, history-aware adaptive ROM algorithm \cite{hedayat_history-aware_2026} and self-adaptive ROM algorithm
\cite{mohaghegh_self_2026}, submitted in peer-reviewed venues. In addition, CompFlowLab has also been employed to generate simulation datasets for 1D RDE
configurations and related reacting-flow problems, providing a computationally tractable platform for investigating data-driven modeling approaches involving coupled shock dynamics and chemical reactions.

\emph{Primary user groups:} CompFlowLab is primarily intended for researchers in model
reduction, scientific machine learning (SciML), and CFD, who require physically representative yet computationally manageable benchmark problems. For the model-reduction and SciML community, it provides direct access to governing-equation operators and simulation data needed to develop and compare intrusive and non-intrusive methodologies. For CFD researchers, its standalone FOM and test cases offer an accessible environment for investigating new numerical methods for compressible flows, including shock-capturing schemes, temporal integration, and physical models for reacting-flow problems. In addition, its compact Python implementation and unified interface also make it suitable for educational applications, allowing students to explore numerical simulation, reduced-order modeling, and data-driven techniques within a common software environment. 
code base.

\begin{sloppypar}
\emph{Reusable software components:} Beyond its integrated simulation capabilities, CompFlowLab provides several modular components that can be reused or extended independently, which include: (1) online basis-update algorithms in
\texttt{rom/basis\_func.py}, (2) hyper-reduction sampling strategies in
\texttt{rom/sampling\_func.py}, (3) the Cantera coupled Roe flux formulation for
multi-species compressible flows, and (4) the simplified 1D RDE injection model with precomputed injection states. These components are available under the permissive MIT license, facilitating their adaptation and integration into other research software.
\end{sloppypar}

\emph{Maintenance and community development:} CompFlowLab is actively maintained and distributed as open-source software under the MIT license. Comprehensive documentation, including installation intructions, software architecture, and benchmark configurations, is available at \url{https://compflowlab.mintlify.app}. Ongoing development focuses on expanding the supported modeling capabilities, improving usability, and facilitating contributions from the broader research community. Through its accessible implementation, modular design, and standardized benchmark environment, CompFlowLab aims to support reproducible research and accelerate the development of advanced data-driven modeling methodologies for compressible flows. 

\section{Conclusions}
\label{sec:conclusions}

This paper presents CompFlowLab, an open-source Python-based software package designed to facilitate developing, prototyping, and systematically evaluating data-driven modeling methodologies for compressible flows. It integrates a 1D finite-volume solver for the Navier--Stokes equations, with the modeling capabilities of handling detailed chemical kinetics through Cantera, and a suite of reduced-order modeling (ROM) approaches, including conventional projection-based ROMs, quadratic-manifold ROMs,
learned closure formulations, adaptive and self-adaptive ROMs.
Multiple interchangeable hyper-reduction sample strategies and online basis-update algorithms further enable flexible development and evaluation of advanced ROM methodologies. Through a modular architecture built around shared data structures and standardized interfaces, CompFlowLab allows new modeling approaches to be incorporated with minimal modifications to the existing codebase. Its unified input configuration supports FOM data generation, ROM construction, and comparative evaluation using consistent governing equations and numerical discretizations, while providing access to intermediate solution data and numerical operators for detailed analysis. Having already supported research on advanced ROM techniques, CompFlowLab provides an accessible and extensible benchmark environment for the model-reduction, scientific machine-learning, and CFD communities, particularly for convection-dominated flows involving shocks, flames, and detonations. 

\subsection*{Limitations}
\begin{sloppypar}
Despite these capabilities, several limitations remain. First, CompFlowLab is fully vectorized with NumPy, and therefore the wall-clock acceleration achieved by ROM over FOM can be limited, especially for relatively small 1D benchmark problems considered here. For assessing the computational savings associated with hyper-reduction, CompFlowLab also provides a loop-based non-vectorized implementation (\texttt{numpy\_vector = False}), in which the cost of residual evaluation more directly scales with the number of cells processed. This option facilitates investigations of algorithmic cost reduction, although the resulting speedups should not be interpreted as representative of performance relative to an optimized vectorized FOM. Second, CompFlowLab assumes a basic understanding of CFD and model-reduction methods, and inappropriate numerical or physical configurations may lead to simulation instability or failure without sufficiently informative diagnostic messages. Third, implicit time integration is currently limited to backward Euler method, and advanced implicit methods, such as higher-order BDF schemes, preconditioning, precise analytical Jacobians, are not yet available. Finally, the current implementation is restricted to serial CPU execution. Computationally demanding reacting-flow cases such as 1D RDE require very small time-step size and hundreds of thousands of time steps, thus several hours of wall-clock time, limiting the efficiency of extensive parametric investigations.
\end{sloppypar}

\subsection*{Future work}

Based on the identified limitations, future development will focus on improving software robustness, computational performance, and modeling capabilities. Enhanced input validation and diagnostic reporting will be introduced to help users identify incompatible configurations, physically inadmissible states, and numerical instabilities. The time-integration module will be extended to support more advanced implicit schemes, including higher-order backward differentiation formulas and improved nonlinear solution strategies.  Paralell computing capabilities, potentially through MPI/GPU acceleration, will be explored to reduce the computational cost of long-duration reacting-flow simulations. In addition, ongoing efforts seek to incorporate data-assimilation-enabled ROMs that correct the model predictions with sparse observations. Prototype solvers based on the ensemble Kalman filter \cite{evensen_sequential_1994} and on Gaussian-mixture-model filtering \cite{sondergaard_data_2013} are included in the repository as experimental modules (\texttt{DACAROM}, \texttt{GMMDOROM}). Further development and validation of these methods will be a priority for future releases, extending CompFlowLab toward a more comprehensive platform for physics-based and data-informed reduced-order modeling.

\section*{Acknowledgements}
\label{}

The authors acknowledge the supports from the Air Force Office of Scientific Research (AFOSR) the grant FA9550- 23-1-0211 (Program managers: Drs. Chiping Li and Fariba Fahroo). Ali Mohaghegh also acknowledges the support from the Sloan Foundation via award
G-2022-19347 through the US Research Software Sustainability Institute (URSSI) Early Career Fellowship program.

\section*{Declaration of generative AI and AI-assisted technologies}
During the preparation of this work, the authors used generative AI for debugging code and manuscript editing. The authors reviewed and edited the output as needed and take full responsibility for the content of the published article. 






\bibliographystyle{unsrt}
\bibliography{references}


\end{document}